\documentclass[11pt,onecolumn,twoside]{IEEEtran}

\usepackage{cite}      
\usepackage{graphicx}  
\usepackage{amsmath}   
\usepackage{amssymb}   

\newtheorem{thm}{Theorem}
\newtheorem{cor}{Corollary}
\newtheorem{lem}{Lemma}
\newtheorem{prop}{Proposition}
\newtheorem{defn}{Definition}

\def\Rreuse{R_{\mathrm{reuse}}}
\def\Robsolete{R_{\mathrm{obsolete}}}
\def\Rupdate{R_{\mathrm{update}}}
\def\Rold{R_{\mathrm{old}}}
\def\Rnew{R_{\mathrm{new}}}

\title{Malleable Coding with Fixed Reuse}%
\author{Lav R.~Varshney,  
   	  Julius Kusuma, 
        and Vivek K~Goyal%
\thanks{This work was supported in part by the NSF Graduate
	  Research Fellowship, Grant CCR-0325774, and Grant CCF-0729069.}%
\thanks{L.~R.~Varshney was with the Department of Electrical Engineering and Computer Science, 
        the Research Laboratory of Electronics, and the Laboratory for Information and Decision Systems,
	  Massachusetts Institute of Technology, Cambridge, MA 02139 USA.  He is now with the
	  IBM Thomas J.\ Watson Research Center, Hawthorne, NY 10532 USA (e-mail: varshney@alum.mit.edu).}%
\thanks{J.~Kusuma is with Schlumberger-Doll Research Center,
        Cambridge, MA 02139 USA (e-mail: kusuma@alum.mit.edu).}%
\thanks{V.~K.\ Goyal is with the Department of Electrical Engineering and Computer Science and
        the Research Laboratory of Electronics,
        Massachusetts Institute of Technology,
        Cambridge, MA 02139 USA (e-mail: vgoyal@mit.edu).}%
}

\begin{document}

\maketitle

\begin{abstract}
In cloud computing, storage area networks, remote backup storage, and similar 
settings, stored data is modified with updates from new versions.  
Representing information and modifying the representation are both expensive. 
Therefore it is desirable for the data to not only be compressed but to also be 
easily modified during updates.
A malleable coding scheme considers both compression efficiency and 
ease of alteration, promoting codeword reuse.  We examine the 
trade-off between compression efficiency and malleability cost---the 
difficulty of synchronizing compressed versions---measured as the length of a 
reused prefix portion.  Through a coding theorem, the region of achievable rates and 
malleability is expressed as a single-letter optimization.
Relationships to common information problems are also described.
\end{abstract}


\section{Introduction}
\label{sec:introduction}

\IEEEPARstart{C}{onventional} data compression uses a small number of compressed-domain symbols but otherwise picks
the symbols without care. This carelessness renders codewords utterly disposable; little can be salvaged
when the source data changes even slightly. Such data compression is concerned only with reducing the
length of coded representations.
Associating costs with changes to the coded representations introduces new trade-offs
and inspires the adoption of a green-age mantra:
\emph{reduce}, \emph{reuse}, \emph{recycle}. 

As an abstraction of several scenarios,
suppose that after compressing a random source sequence $X_1^n$, it is modified to 
become a new source sequence $Y_1^n$ according to an
update process $p_{Y|X}$.
A \emph{malleable coding} scheme preserves 
a portion of the codeword of $X_1^n$ and modifies the remainder
into a new codeword from which $Y_1^n$ may be 
decoded reliably.

There are several possible notions of preserving a portion of a
codeword.  Here we consider reusing a fixed part of the codeword for $X_1^n$ 
in generating a codeword for $Y_1^n$.
We call this \emph{fixed reuse} since a segment is cut
from the old codeword and reused as part of the new codeword.
Without loss of generality, the fixed portion can be taken
to be at the beginning, so the new codeword is a fixed
prefix followed by a new suffix.  

The fixed reuse formulation is suitable for applications where the update information (new suffix) must
be transmitted through a rate-limited communication channel. If the locations of changed symbols were
arbitrary, the locations would also need to be communicated, communication which may be prohibitively
costly. A contrasting scenario is for a cost to be incurred when a symbol is changed
in value, regardless of its location.  We studied this random access problem in \cite{VarshneyKG2009}.

Our main result is a characterization of achievable rates as a
single-letter expression.
To the best of our knowledge, this is among the first works connecting problems
of information storage---communication across time---with problems in
multiterminal information theory. 
In particular, a connection to the G\'{a}cs--K\"{o}rner 
common information shows that a large
malleability cost must be incurred if the rates for the two versions are required 
to be near entropy.

The remainder of the paper is organized as follows.
Section~\ref{sec:background} gives engineering motivation and 
Section~\ref{sec:problemstatement} provides a formal problem statement.  
The region describing the trade-off
between the rates for the original codeword, for the reused portion,
and for the new codeword is the main object of study.
Section~\ref{sec:MarkovSimple} uses an implicit Markov property to simplify the
analysis of the rate region and Section~\ref{sec:twopoints} describes two easily achieved points.
Theorem~\ref{thm:opt1} in Section~\ref{sec:cutpaste} gives 
the achievable rate region in terms of an auxiliary random variable.  
Section~\ref{sec:auxchar} looks at the auxiliary 
random variable in detail.  
Section~\ref{sec:connections} connects this malleable coding problem to 
other problems in multiterminal information theory.
Section~\ref{sec:final} closes the paper.

\section{Technological Motivations}
\label{sec:background}
Our study of malleable coding is primarily motivated by several
kinds of information technology infrastructures where there is a separation
between terminals used to process information and storage devices 
used to store information.
Many such systems store frequently-updated documents having 
versions whose contents differ only slightly  
\cite{BobbarjungJD2006,PolicroniadesP2004,BurnsSL2003,SuelM2003}.
Moreover, old versions need not be preserved.
Correlations among versions differentiates malleable coding from 
write-efficient memories \cite{AhlswedeZ1989}, where
messages are assumed independent.

Storage area network (SAN) and network-attached storage (NAS) systems comprise a communication infrastructure 
for physical connections and a management infrastructure for organizing connections, 
storage elements, and computers for robust and efficient data transfers
\cite{Lala2003,Jepsen2004}.  Grid computing and distributed storage systems
have similar distributed caching \cite{WangWHPGDK2007,DimakisR2008},
as do cloud computing systems where the complicated interplay between storage and 
transmission is even further enhanced \cite{ArmbrustFGJKKLPRSZ2009,RossRJBCMSMUSJGFPZ2011}.
Even within single computers, updating caches within the 
memory hierarchy involves data transfers among levels \cite{PattersonH1998}.  

Current technological trends in transmission and storage technologies
show that transmission capacity has grown more slowly than
disk storage capacity \cite{ArmbrustFGJKKLPRSZ2009,WangWHPGDK2007}.  Hence ``new''
representation symbols may be more expensive than ``old''
representation symbols, suggesting that reusing parts of codewords 
may be more economical than simply reducing their lengths,
as in conventional data compression.

In cloud computing, cost and latency differentials between storage and transmission of data lead
to data transfer bottlenecks, though as noted, ``once data is in the
cloud for any reason it may no longer be a bottleneck'' \cite{ArmbrustFGJKKLPRSZ2009}.  
Reusing stored data may be of significant value for this emerging technology.

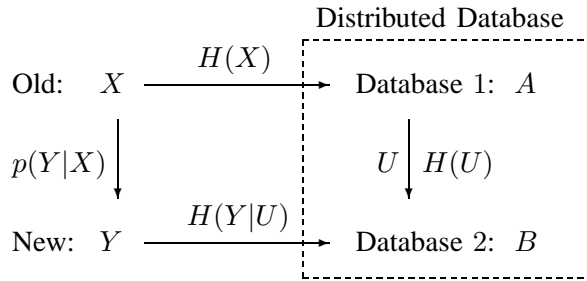
\begin{figure}
  \vspace{1.5ex}
  \centering
  \setlength{\unitlength}{10pt}
  \begin{picture}(24.0,10.0)
  \put(1.0,7.0){Old:}
  \put(1.0,1.0){New:}
  \put(5.0,6.0){\vector(0,-1){3.0}}
  \put(1.0,4.0){$p(Y|X)$}
  \put(4.3,7.0){$X$}
  \put(4.3,1.0){$Y$}
  \put(8.0,8.0){$H(X)$}
  \put(6.0,7.3){\vector(1,0){7.0}}
  \put(7.6,2.0){$H(Y|U)$}
  \put(6.0,1.3){\vector(1,0){7.0}}
  \put(12.0,0.0){\dashbox{0.2}(11.0,9.0)}
  \put(12.5,9.5){Distributed Database}
  \put(14.0,7.0){Database $1$:}
  \put(14.0,1.0){Database $2$:}
  \put(20.0,7.0){$A$}
  \put(20.0,1.0){$B$}
  \put(16.0,6.0){\vector(0,-1){3.0}}
  \put(16.5,4.0){$H(U)$}
  \put(14.8,4.0){$U$}
  \end{picture}
  \caption{Distributed database access.}
  \label{fig:system}
\end{figure}

For several concrete scenarios,
consider the topology given in Fig.~\ref{fig:system}.  The first
user has stored a codeword $A$ for document $X$ in database
$1$. Now the second user, who has a copy of $X$, modifies it to 
create $Y$.  The second user wants to save the new
version to the information system, but since the users are separated, 
database $2$ rather than database $1$ serves this user.  Transmission 
costs for different links may be different.  The natural problem
is to minimize the total cost needed to create a codeword $B$ at 
database $2$ that losslessly represents $Y$.

Consider two users who both have
access to a distributed database system that stores several
copies of the first user's document on different media
at different locations.  Due to proximity considerations, the users will access 
the document from different physical stores.  Suppose that
the first user downloads and edits her document and then wishes to send 
the new version to the second user.  There are different
ways to accomplish this.  The first user can send the entire new version to
the second user or the second user can download the old version from his local store
and require that the first user only send the modification.  
In the former scheme, the cost of transmission is borne entirely by the 
links between the users, rendering distributed storage pointless.
In the latter scheme, there is a trade-off between the rate 
at which the second user downloads the 
original version from the database system and the rate at
which the first user communicates the modification.

Even in a single user scenario, there may be similar considerations.
The first user may simply wish to update the storage device with 
her edited version.  The goal would be to avoid having to create an
entirely new version of the stored codeword by taking advantage of the
availability of the stored original in the database.  

Finally, recent advances in biotechnology have demonstrated storage 
of artificial messages in the DNA of living organisms \cite{WongWF2003}. 
Such systems provide another motivating application, since
certain biotechnical editing costs correspond to the malleability costs defined
for fixed reuse.

\section{Problem Statement}
\label{sec:problemstatement}

Let $\{(X_i,Y_i)\}_{i=1}^{\infty}$ be a sequence of independent drawings
of a pair of random variables $(X,Y)$, $X \in \mathcal{W}$, $Y \in \mathcal{W}$, where
$\mathcal{W}$ is a finite set and $p_{X,Y}(x,y) = \Pr[X = x, Y=y]$.
The joint distribution determines the marginals, $p_X(x)$ and 
$p_Y(y)$, as well as the modification channel, $p_{Y|X}(y|x)$.
Denote the storage medium alphabet by $\mathcal{V}$, which is also a finite set.
It is natural to measure all rates in numbers of symbols from $\mathcal{V}$.
This is analogous to using base-$|\mathcal{V}|$ logarithms, and all logarithms 
should be so interpreted.

\begin{figure}
  \vspace{1.5ex}
  \centering
  \includegraphics[width=3.5in]{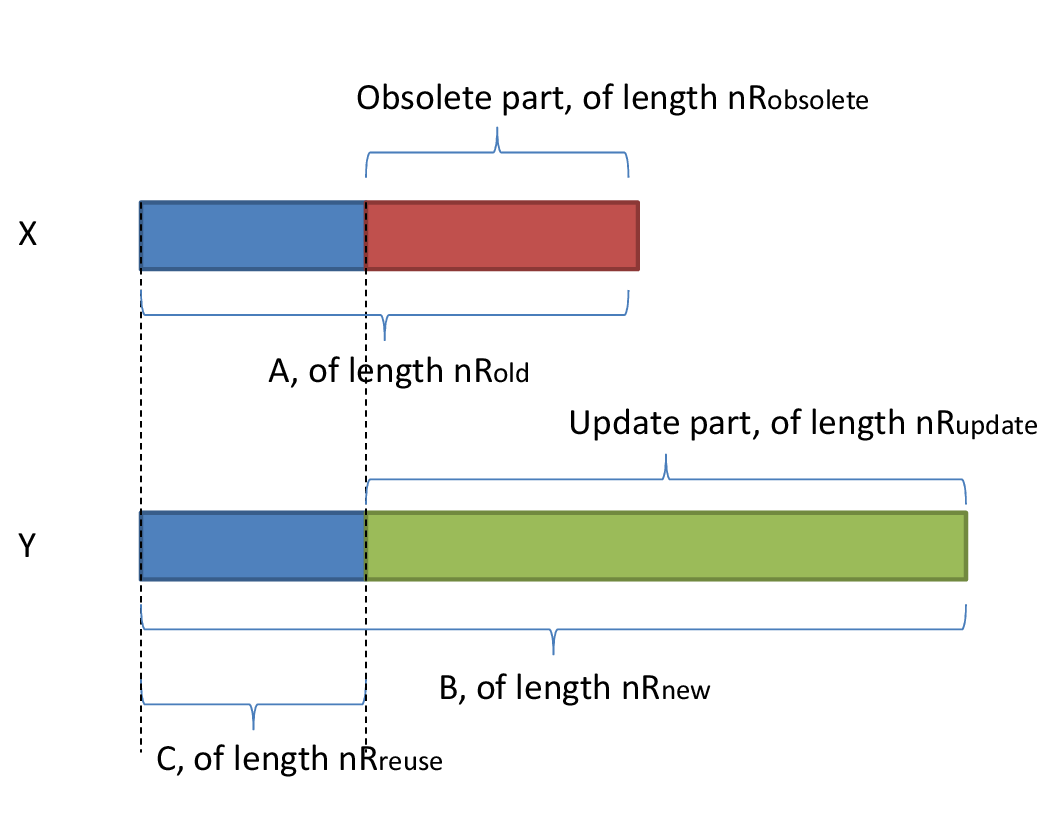}
  \caption{In malleable coding with fixed reuse, the compressed
    representations of $X_1^n$ and $Y_1^n$ have the first $n\Rreuse$ storage symbols
    in common.}
  \label{fig:compressed}
\end{figure}

Our interest is in coding of $X_1^n$ followed by coding of $Y_1^n$
where the first $n\Rreuse$ letters of the codewords are exactly the same.
As depicted in Fig.~\ref{fig:compressed},
$A_1^{n\Rold} \in \mathcal{V}^{n\Rold}$ is the representation of $X_1^n$,
$B_1^{n\Rnew} \in \mathcal{V}^{n\Rnew}$ is the representation of $Y_1^n$,
and $C_1^{n\Rreuse} \in \mathcal{V}^{n\Rreuse}$ is the common part.
The parts not in common are of lengths $n\Robsolete$ and $n\Rupdate$ respectively.
Encoder and decoder mappings are thus defined as follows.

An encoder for $X$ with parameters $(n,\Rreuse,\Rold)$ is the concatenation of two mappings:
\[
f_E^{(X)} = f_E^{(U)} \times f_E^{\prime(X)} \mbox{,}
\]
where
\[
f_E^{(U)}: \mathcal{W}^n \to \mathcal{V}^{n\Rreuse} \mbox{ and } 
f_E^{\prime(X)}: \mathcal{W}^n \to \mathcal{V}^{n\Robsolete} \mbox{.}
\]
An encoder for $Y$ with parameters $(n,\Rreuse,\Rnew)$ is defined as:
\[
f_E^{(Y)} = f_E^{(U)} \times f_E^{\prime(Y)} \mbox{,}
\]
where 
we use one of the previous encoders $f_E^{(U)}$ together with 
\[
f_E^{\prime(Y)}: \mathcal{W}^n \times \mathcal{V}^{n\Rreuse} \to \mathcal{V}^{n\Rupdate} \mbox{.}
\]
Notice that $f_E^{\prime(Y)}$ is defined so as to have access to the previously stored prefix.
Given these encoders, a common decoder with parameter $n$ is
\[
f_D: \mathcal{V}^{*} \to \mathcal{W}^n = \begin{cases}\mathcal{V}^{n\Rold} \to \mathcal{W}^n\mbox{,} & \mbox{first version}\\ \mathcal{V}^{n\Rnew} \to \mathcal{W}^n\mbox{,} & \mbox{second version.}\end{cases}
\]
The encoders and decoder define a block code for fixed reuse malleability.  

A trio $(f_E^{(X)},f_E^{(Y)},f_D)$ with parameters $(n,\Rreuse,\Rold,\Rnew)$
is applied as follows.  Let 
\[
A_1^{n\Rold} = f_E^{(X)}(X_1^n) = [f_E^{(U)}(X_1^n),f_E^{\prime(X)}(X_1^n)] \mbox{,}
\]
$A_1^{n\Rold} \in \mathcal{V}^{n\Rold}$, be the source code for $X_1^n$, where the first part of the code---which
will be reused---is explicitly
notated as 
\[
C_1^{n\Rreuse} \in \mathcal{V}^{n\Rreuse}= f_E^{(U)}(X_1^n)\mbox{.}
\]
The partial codeword $C_1^{n\Rreuse}$ asymptotically almost surely (a.a.s.) losslessly represents a random variable we call $U_1^n$.  
Then the encoding of $Y_1^n$ is carried out as 
\begin{eqnarray*}
B_1^{n\Rnew} & = & f_E^{(Y)}(C_1^{n\Rreuse},Y_1^n) \\
           & = & [C_1^{n\Rreuse}, f_E^{\prime(Y)}(C_1^{n\Rreuse},Y_1^n)]\mbox{,}
\end{eqnarray*}
$B_1^{n\Rnew} \in \mathcal{V}^{n\Rnew}$.
We also let 
\[
(\hat{X}_1^n,\hat{Y}_1^n) = (f_D(A_1^{n\Rold}),f_D(B_1^{n\Rnew})) \mbox{.}
\]
We define the error rate
\[
\Delta = \max(\Delta_X,\Delta_Y) \mbox{,}
\]
where
\[
\Delta_X = \Pr[X_1^n \neq \hat{X}_1^n] \mbox{ and }
\Delta_Y = \Pr[Y_1^n \neq \hat{Y}_1^n] \mbox{.}
\]
Note that by construction we insist that the first $n\Rreuse$ symbols are identical:
\[
A_1^{n\Rreuse} = B_1^{n\Rreuse} = C_1^{n\Rreuse} \mbox{.}
\]

We use conventional performance criteria for the code, which are the numbers of
storage-medium letters per source letter
\[
\Rold = \frac{1}{n}\log_{|\mathcal{V}|} |\mathcal{V}|^{n\Rold} \mbox{ and }
\Rnew = \frac{1}{n}\log_{|\mathcal{V}|} |\mathcal{V}|^{n\Rnew} \mbox{,}
\]
and add, as a third performance criterion, the normalized length of the portion of
the code that does not overlap
\[
\Rupdate = \Rnew-\Rreuse = \frac{1}{n}\log_{|\mathcal{V}|} |\mathcal{V}|^{n\Rupdate} \mbox{.}
\]

\begin{defn}
Given a source $p(X,Y)$, a triple $(\Rold^0,\Rnew^0,\Rupdate^0)$ 
is said to be \emph{achievable} if, for arbitrary $\epsilon > 0$, 
there exists (for $n$ sufficiently large) a block code 
with error rate $\Delta \le \epsilon$,
and lengths
$\Rold \le \Rold^0 + \epsilon$, 
$\Rnew \le \Rnew^0 + \epsilon$, and
$\Rupdate \le \Rupdate^0 + \epsilon$.
\end{defn}

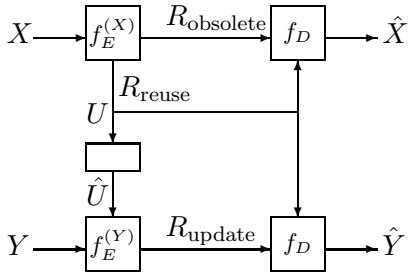
\begin{figure}
  \centering
  \vspace{1ex}
  \setlength{\unitlength}{10pt}
  \begin{picture}(14.0,8.0)(0.0,1.4)

  \put(0.0,9.0){$X$}
  \put(0.0,1.0){$Y$}
  \put(1.0,9.2){\vector(1,0){2.0}}
  \put(1.0,1.2){\vector(1,0){2.0}}
  \put(3.0,8.4){\framebox(2.0,2.0){\footnotesize $f_E^{(X)}$}}
  \put(3.0,0.4){\framebox(2.0,2.0){\footnotesize $f_E^{(Y)}$}}
  \put(5.0,9.2){\vector(1,0){5.0}}
  \put(5.0,1.2){\vector(1,0){5.0}}
  \put(6.0,9.7){$\Robsolete$}
  \put(6.0,1.7){$\Rupdate$}
  \put(10.0,8.4){\framebox(2.0,2.0){\footnotesize $f_D$}}
  \put(10.0,0.4){\framebox(2.0,2.0){\footnotesize $f_D$}}
  \put(12.0,9.2){\vector(1,0){2.0}}
  \put(12.0,1.2){\vector(1,0){2.0}}
  \put(14.2,9.0){$\hat{X}$}
  \put(14.2,1.0){$\hat{Y}$}
  \put(4.0,8.4){\vector(0,-1){3.2}}
  \put(3.0,4.2){\framebox(2.0,1.0){$\quad$}}
  \put(4.0,4.2){\vector(0,-1){1.8}}
  \put(3.0,2.9){$\hat{U}$}

  \put(3.0,6.0){$U$}
  \put(4.2,7.0){$\Rreuse$}
  \put(4.0,6.4){\line(1,0){7.0}}
  \put(11.0,6.4){\vector(0,1){2.0}}
  \put(11.0,6.4){\vector(0,-1){4.0}}

  \end{picture}
  
  \caption{Block diagram for malleable coding with fixed reuse.}
  \label{fig:system1} 
\end{figure}

We want to determine the set of achievable rate triples, $\mathcal{M}$.  
It follows from the definition that $\mathcal{M}$ is a closed subset of
$\mathbb{R}^3$ and has the property that if $(\Rold^0,\Rnew^0,\Rupdate^0) \in \mathcal{M}$,
then $(\Rold^0+\delta_0,\Rnew^0+\delta_1,\Rupdate^0+\delta_2) \in \mathcal{M}$ for any
$\delta_i \ge 0$, $i = 0,1,2$. 
The rate region $\mathcal{M}$ is thus completely defined by its 
lower boundary, which is itself closed.
The triple $(\Robsolete,\Rupdate,\Rreuse)$ may be used in place of $(\Rold,\Rnew,\Rupdate)$ when convenient,
as depicted in Fig.~\ref{fig:system1}.  

\section{Time Ordering, Markov Relations, and Two Achievable Points}

We begin by considering the effect of time ordering on our problem
and give two achievable points. We will later continue with a 
general characterization of the rate region.

\subsection{Simplification}
\label{sec:MarkovSimple}
There is a time ordering in malleable coding. The sources $X_1^n$ and $Y_1^n$
come from a joint distribution, however the partial codeword $C_1^{n\Rreuse}$ that represents 
$U_1^n$ is generated by encoder $f_E^{(U)}$ based on $X_1^n$ prior to the encoding 
of $Y_1^n$ by $f_E^{\prime(Y)}$. Consequently the time ordering of the encoding procedure implies the Markov relation $U \leftrightarrow X \leftrightarrow Y$.  

One might think that expending $\Rold$ greater than $H(X)$ might allow a better 
side information random variable $U_1^n$ to be formed, but expanding the representation
of $X_1^n$ beyond entropy provides no advantage.  That is, any extra bits used to 
encode $X_1^n$ will not help in representing $Y_1^n$.
\begin{prop}
Taking $\Rold > H(X)$ provides no advantage in malleable coding with fixed reuse.
\end{prop}
\begin{IEEEproof}
Consider the representation of $X_1^n$, $A_1^{n\Rold} = [f_E^{(U)}(X_1^n), f_E^{\prime(X)}(X_1^n)]$
and for convenience, let $A_1^{\prime n\Robsolete} = f_E^{\prime(X)}(X_1^n)$ denote the portion that is not reused, so that $A_1^{n\Rold} = [C_1^{n\Rreuse}, A_1^{\prime n\Robsolete}]$.  Suppose we expand the 
representation by taking $\Rold > H(X)$.  The extra symbols are either spent in $C$, in $A^{\prime}$, or in both.

From the time-ordering derived Markov structure, $U\leftrightarrow X \leftrightarrow Y$,
$X$ is a sufficient statistic of $U$ for $Y$.

Spending extra symbols in $A^{\prime}$ is wasteful since $A^{\prime}$ is not used to 
encode $Y_1^n$.  Spending extra symbols in $C_1^{n\Rreuse}$ means that $\Rreuse > H(f_E^{(U)}(X_1^n))$; spending extra symbols in $C_1^{n\Rreuse}$ is wasteful since $X$ is a sufficient
statistic of $U$ for $Y$.
\end{IEEEproof}

We focus on expanding $\Rnew$ beyond $H(Y)$ and analyze the achievable rate region.
Moreso, we focus on how $\Rnew$ depends on the size of the portion to be reused, $\Rreuse$.
In particular,
we fix $\Rreuse$ and find the best $\Rnew$; the smallest $\Rnew$ is denoted $\Rnew^{*}(\Rreuse)$ or alternatively the smallest malleability rate $\Rupdate$ is denoted $\Rupdate^{*}(\Rreuse)$.

\subsection{Two Achievable Points}
\label{sec:twopoints}

It is easy to note the 
values of the corner points corresponding to $\Rreuse = 0$
and $\Rreuse = H(X)$.  
For $\Rreuse=0$, the lossless source coding theorem
yields $\Rnew^*(0) = H(Y)$.  For $\Rreuse = H(X)$, since the lossless
compression of $X_1^n$ has to be preserved, 
$\Rnew^*(H(X)) = H(X,Y)$.  This follows since
the first $H(X)$ symbols are fixed, we need to 
losslessly represent the conditionally typical set, which requires 
$H(Y|X)$ additional symbols, for a total of $H(X)+H(Y|X)= H(X,Y)$. 
Since $H(Y|X) \leq H(Y)$, this is better than discarding the old codeword 
and creating an entirely new codeword for $Y_1^n$; unless $X$ and $Y$ 
are independent, this is strictly better.

\section{Main Results}
\label{sec:cutpaste}

We cast the fixed reuse malleable coding problem as a single-letter information-theoretic optimization.
Unfortunately this is not computable in general, but in the next section 
we will give a computable partial characterization 
for cases where there is a suitable sufficient statistic.

A proof of the Slepian-Wolf distributed source coding theorem uses the method of binning 
\cite{Cover1975, CsiszarK1997}, in which the codebooks for the sources are segmented and 
codewords are binned.  Results are obtained by choosing appropriate bin sizes: 
for two sources, the bin sizes are limited by the mutual information between them.  
However, this approach says nothing about whether or how labels are kept synchronized 
between the different codebooks and bins.  We apply a similar binning approach to the 
codeword labels in the codebooks, but insist on consistent representation to enforce 
malleability in the representations.

We consider the trade-off between $\Rnew$ and $\Rreuse$ (and thus $\Rupdate$). From the previous section, it is clear 
that for a given malleability, the compression efficiency of $Y_1^n$ is determined by 
the quality of the binning in the codebook for $X_1^n$. 
We insist that $U$ is a deterministic function of $X$, i.e., $U = f(X)$. Then,
we can formulate the following information-theoretic optimization problem:
\begin{eqnarray}
\label{eqn:opt1}
\Rupdate^*(\Rreuse)
  & = & \Rnew^*(\Rreuse) - \Rreuse \nonumber \\
  & = & \min\limits_{U:U = f(X), H(U) = \Rreuse} H(Y|U) \mbox{.}
\end{eqnarray}

\begin{thm}
\label{thm:opt1}
The optimization problem \eqref{eqn:opt1} provides a boundary 
to the rate region $\mathcal{M}$ when $\Rold = H(X)$.
\end{thm}

For clarity, before stating the proof to Theorem~\ref{thm:opt1} we describe the dimensions 
and alphabets of the codebooks used.
\begin{enumerate}
  \item Numbers $\Rreuse$ and $\Rold$ are given.  The first codebook is used to encode a source sequence
	of length $n$, $x_1^n$.  Let $\mathcal{C} = \{c_1,c_2,\ldots,c_{\rho_u}\}$ be the prefix-stage
	codebook of size $\rho_u = \lceil |\mathcal{V}|^{n\Rreuse}\rceil$, drawn from the
	alphabet $\mathcal{V}$.  Corresponding to every codeword $c_i \in \mathcal{C}$, let 
	$\mathcal{A}^{\prime}(c_i) = \{a_1(c_i),a_2(c_i),\ldots,a_{\rho_{x^{\prime}}}(c_i)\}$ be
	the suffix-stage codebook of size $\rho_{x^{\prime}} = \lceil |\mathcal{V}|^{n\Robsolete} \rceil$, 
	drawn from the alphabet $\mathcal{V}$.  The whole codebook for $x_1^n$ is then
	$\mathcal{A} = \cup_{i=1}^{n\Rreuse} \mathcal{A}^{\prime}(c_i)$ which is a tree-structured codebook
	of size $\lceil |\mathcal{V}|^{n\Rold} \rceil$.
  \item The prefix-stage codebook $\mathcal{C}$ from above and a number $\Rnew$ is given.  The second 
	codebook is used to encode a source sequence of length $n$, $y_1^n$.  Corresponding to every
	codeword $c_i \in \mathcal{C}$, let 
	$\mathcal{B}^{\prime}(c_i) = \{b_1(c_i),b_2)c_i),\ldots,b_{\rho_{y^{\prime}}}(c_i)\}$ be the
	suffix-stage codebook of size $\rho_{y^{\prime}} = \lceil |\mathcal{V}|^{n\Rupdate} \rceil$, 
	drawn from alphabet $\mathcal{V}$.  The whole codebook for $y_1^n$ is then
	$\mathcal{B} = \cup_{i=1}^{n\Rreuse}\mathcal{B}^{\prime}(c_i)$ which is a tree-structured codebook
	of size $\lceil |\mathcal{V}|^{n\Rnew} \rceil$.
\end{enumerate}
The two codebooks share the first level of the tree, but have different second levels.

The proof of Theorem~\ref{thm:opt1} makes use of the following lemma due to K\"{o}rner \cite{Korner1971}.
\begin{lem}[\cite{Korner1971}]
\label{lem:Korner}
Let $\{\xi_i\}_{i=1}^{\infty}$ be a discrete, memoryless source
drawn from the finite alphabet $\mathcal{W}$.  Let $f$ be a function on $\mathcal{W}$
that partitions $\mathcal{W}$.  For $a, b \in \mathcal{W}$, let $a|b$
denote the condition $f(a) = f(b)$ and $a \neq b$.  For a set $A \subset \mathcal{W}^n$, let
\begin{align*}
[A] &= \min \{r: \ A = \cup_{i=1}^{r} A_i,\ A_i \cap A_j = \emptyset \mbox{ for } i \neq j \\
&\qquad\qquad\mbox{ and } a,b \in A_i \Rightarrow a|b \ \mbox{does not hold} \}
\end{align*}
Let
\[
M(n,\lambda) = \min_{A \subset \mathcal{W}^n: \Pr[\xi_1,\xi_2,\ldots,\xi_n \in A]\ge 1-\lambda} [A]
\]
Then for every $\lambda$, $0 \le \lambda < 1$, $\lim_{n\to\infty} \tfrac{1}{n} \log_2 M(n,\lambda)$ exists and satisfies
\[
\lim_{n\to\infty} \tfrac{1}{n} \log_2 M(n,\lambda) = H(\xi | f(\xi))\mbox{.}
\]
\end{lem}
This lemma concerns itself with the smallest partition of a set $A$ that
allows one to almost surely disambiguate the set partitions of $A$ given
that one observes a function of members of these partitions.
K\"{o}rner's result states that for any function $f$ that partitions the
alphabet $\mathcal{W}$, the minimum rate required to disambiguate $\xi$ if
the decoder has side information $f(\xi)$ is $H(\xi | f(\xi))$.

We now state the proof to Theorem~\ref{thm:opt1}.

\begin{IEEEproof}
Fix a function $f$ that partitions $\mathcal{W}$.  This function is used to induce
a random variable $U_1 = f(X_1)$.  The function $f$ is applied to all $X_1^n$ in the same 
manner to produce the memoryless random variables $U_1^n$.

\paragraph{Generating the first codebook}
Choose the prefix part codebook rate as $\Rreuse = \tfrac{1}{n} \log_{|\mathcal{V}|}\rho_u = H(U) + \delta_1(n)$,
where $\delta_1(n) \to 0$ as $n\to\infty$.
Generate a set of size $|\mathcal{V}|^{n\Rreuse}$ of sequences in $\mathcal{W}^n$
with elements drawn i.i.d.\ according to $p_{U}$.  Now take these sequences in order and
create a codebook $\mathcal{C}$ with codewords from $\mathcal{V}^{n\Rreuse}$ listed in lexicographic
order, by making a one-to-one correspondence between the two sets (which are of the same 
size).\footnote{Note that this codebook generation procedure is different than putting the 
typical set of source sequences into correspondence with the codebook, which is common in proofs of 
the source coding theorem. Rather, it is random code generation, which is common
in proofs of the channel coding theorem.}

Use K\"{o}rner's optimal complementary code (the existence of which is promised by 
Lemma~\ref{lem:Korner}) as the suffix-part codebook $\mathcal{A}^{\prime}$. As given in 
Lemma~\ref{lem:Korner}, it should have rate $\Robsolete = \tfrac{1}{n}\log_{|\mathcal{V}|}\rho_{x^{\prime}}
= H(X|f(X)) + \delta_2(n) = H(X|U)+ \delta_2(n)$, where $\delta_2(n) \to 0$ as $n\to\infty$.

Notice that with the choices of $\Rreuse$ and $\Robsolete$ given,
\begin{align*}
\Rold \approx H(U) + H(X|U) &\stackrel{(a)}{=} H(X,U) \\ 
&\stackrel{(b)}{=} H(X)
\end{align*}
where (a) is due to the chain rule of entropy and (b) is due to the fact that $f(\cdot)$ is a 
deterministic function.

The codebook $\mathcal{A} = [\mathcal{C}, \mathcal{A}^{\prime}]$ is revealed to both the 
encoder and decoder.

\paragraph{Encoding the first version}
For a source realization $x_1^n$, compute $u_1^n = f(x_1^n)$.  If $u_1^n$ is represented
in the codebook $\mathcal{C}$, then its corresponding codeword is written to the storage medium 
in the prefix-part position. If $u_1^n$ is not represented in the codebook, then a codeword in 
$\mathcal{C}$ is chosen uniformly at random from $\mathcal{C}$ and written to the storage medium in the prefix-part position.

For the suffix-part position, if $u_1^n$ was represented by $c_{u_1^n} \in \mathcal{C}$
and if $x_1^n$ is represented in the codebook $\mathcal{A}^{\prime}(c_{u_1^n})$, then
its corresponding codeword is written to the storage medium.  If $u_1^n$ was represented by
$c_{u_1^n} \in \mathcal{C}$ and if $x_1^n$ is not represented in the codebook $A^{\prime}(c_{u_1^n})$,
then the all-zeros sequence in $\mathcal{V}^{n\Robsolete}$ is written to the suffix-part position 
of the storage medium. Likewise, if $u_1^n$ was not represented by some
$c_{u_1^n} \in \mathcal{C}$, then the all-zeros sequence in $\mathcal{V}^{n\Robsolete}$ is written to 
the suffix-part position of the storage medium.

\paragraph{Decoding the first version}
Decoding is performed using lookup in $\mathcal{A}$ to generate $\hat{x}_1^n \in \mathcal{W}^n$,
the recovered version of $x_1^n$.

\paragraph{Error analysis for first version}
The two possible error events are the following:
\begin{enumerate}
  \item $\mathcal{E}_1$: $u_1^n$ is not represented in $\mathcal{C}$; and
  \item $\mathcal{E}_2$: $u_1^n$ is represented by $c_{u_1^n} \in \mathcal{C}$, 
	but $x_1^n$ is not represented in $A^{\prime}(c_{u_1^n})$.
\end{enumerate}
The codebook $\mathcal{C}$ represents $|\mathcal{V}|^{n(H(U) + \delta_1(n))}$ sequences generated 
i.i.d.\ according to $p_U$. The probability that a source sequence $u_1^n$ generated i.i.d.\ 
according to $p_U$ is identical to the first codeword of the codebook is bounded as $|\mathcal{W}|^{-n}$,
by memorylessness and the length of the codebook.

Since these identicality events are independent, for a codebook of size $|\mathcal{V}|^{n(H(U) +\delta_1(n))}$, the probability of $\mathcal{E}_1$ is therefore bounded as
\[
\Pr[\mathcal{E}_1] \le 1 - \left[1 - |\mathcal{W}|^{-n} \right]^{|\mathcal{V}|^{n(H(U) +\delta_1(n))}} 
\]
which goes to zero as $n \to \infty$.

Furthermore, Lemma~\ref{lem:Korner} guarantees that $\Pr[\mathcal{E}_2] \to 0$ as $n \to \infty$.
Thus by the union bound, the total error probability goes to zero asymptotically.

\paragraph{Converse arguments for first version}
By the converse of the source coding theorem \cite{Shannon1948}, the size of
$\mathcal{C}$ cannot be chosen smaller than $H(U)$ to drive the error probability to zero 
as $n \to \infty$. By the converse part of Lemma~\ref{lem:Korner}, the suffix-part of the code 
cannot be chosen smaller than $H(X|U)$ to drive the error probability to zero as $n \to \infty$.

\paragraph{Decoding the prefix for use with the second version}
The prefix-part is preserved in its entirety on the storage medium, therefore $c$ 
is identical to above. For a given blocklength $n$, it can be used to decode $u_1^n$
with an error probability $\Pr[\mathcal{E}_1] = \epsilon$, $\epsilon(n) \to 0$ as $n \to \infty$. 
The decoded version is called $\hat{u}_1^n$: note that $\hat{U}_1^n$ is a memoryless sequence
of random variables because the codebook $\mathcal{C}$ is a random codebook with i.i.d.\
$p_U$ entries and since error events lead to a uniformly random choice of codeword within $\mathcal{C}$.

\paragraph{Generating the second codebook} The prefix part has the same codebook $\mathcal{C}$ as above.
For the suffix part, consider generating the codebook according to the memoryless random variable $(Y_1^n,\hat{U}_1^n)$
when the decoder is assumed to have side information $\hat{U}_1^n$. 
Since $g(Y, \hat{U}) = \hat{U}$ is a function that partitions the space, we can
use K\"{o}rner's optimal complementary code (the existence of which is promised by 
Lemma~\ref{lem:Korner}) as the suffix-part code $\mathcal{B}^{\prime}$. 
As given in Lemma~\ref{lem:Korner}, it should have rate $\Rupdate = \tfrac{1}{n} \log_{|\mathcal{V}|} \rho_{y^{\prime}} = H((Y, \hat{U})|\hat{U}) = H(Y|\hat{U})$.

By a continuity argument, Lemma~\ref{lem:continuity} in the appendix, $H(Y|\hat{U}) - H(Y|U)$ goes to 
zero as $n \to \infty$, and so we can take $\Rupdate = H(Y |U)$.

The codebook $\mathcal{B} = [\mathcal{C}, \mathcal{B}^{\prime}]$ is revealed to both the 
encoder and decoder.

\paragraph{Encoding the second version}
The prefix part is as for the first version, $b_1^{n\Rreuse} = c_1^{n\Rreuse}$.

For the suffix-part $b^{n\Rnew}_{n\Rreuse+1}$, let $\hat{u}_1^n$ be represented by $c_{\hat{u}_1^n} \in \mathcal{C}$.
If $y_1^n$ is represented in the codebook $\mathcal{B}^{\prime}(c_{\hat{u}_1^n})$, then its corresponding
codeword is written to the storage medium. If $y_1^n$ is not represented in the codebook 
$\mathcal{B}^{\prime}(c_{\hat{u}_1^n})$, then the all-zeros sequence in $\mathcal{V}^{n\Rupdate}$
is written to the suffix-part position of the storage medium.

\paragraph{Decoding the second version}
Decoding is performed using lookup in $\mathcal{B}$ to generate $\hat{y}_1^n \in \mathcal{W}^n$,
the recovered version of $y_1^n$.

\paragraph{Error analysis for second version}
There is one possible error event:
\begin{enumerate}
  \item $\mathcal{E}_3$: $y_1^n$ is not represented in $\mathcal{B}^{\prime}(c_{\hat{u}_1^n})$.
\end{enumerate}
Lemma~\ref{lem:Korner} guarantees that $\Pr[\mathcal{E}_3] \to 0$ as $n \to \infty$.

\paragraph{Converse arguments for second version} 
By the converse part of Lemma~\ref{lem:Korner}, the suffix-part of the
codebook cannot be chosen smaller than $H(Y|U)$ to drive the error probability to zero as $n \to \infty$.
\end{IEEEproof}

\section{Further Characterizations}
\label{sec:auxchar}

As in the source coding with side information problem 
\cite{AhlswedeK1975,Wyner1975b,MarcoE2009} and elsewhere, 
Theorem~\ref{thm:opt1} left us
to optimize an auxiliary random variable $U$ that describes the method of binning.  
Here we give further characterization in terms of $W$, a minimal 
sufficient statistic of $X$ for $Y$.

\begin{figure}[ht]
  \vspace{1.5ex}
 \centering
  \includegraphics[width=2.6in]{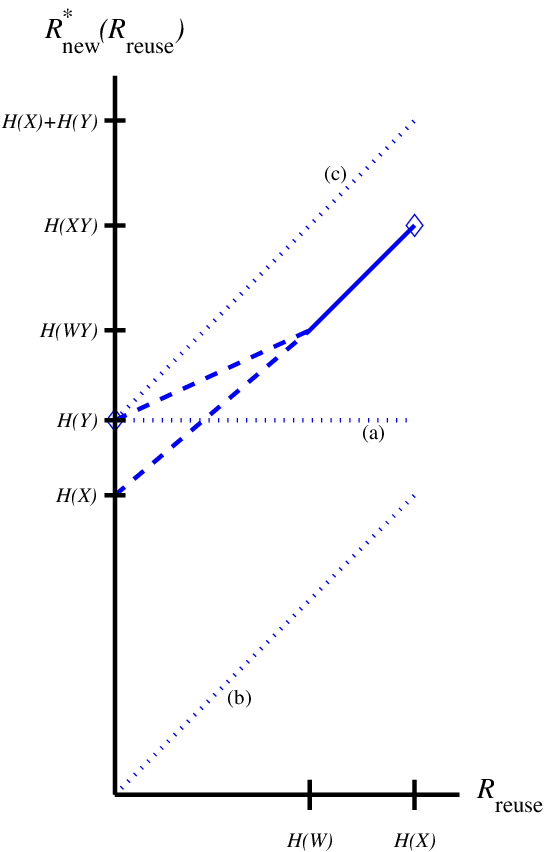}
  \caption{Characterizations of the rate region boundary $\Rnew^*(\Rreuse)$.
    Each $\Diamond$ is a point determined in Section~\ref{sec:twopoints},
    and the dotted lines are simple bounds from Section~\ref{sec:auxchar}.
    With $W$ defined as a minimal sufficient statistic of $X$ for $Y$,
    the solid line shows the unit-slope boundary determined by
    Theorem~\ref{thm:lm_regime2}.
    The dashed lines demarcate the portion of boundary that is unknown
    (but known to be convex by Theorem~\ref{thm:lm_regime1}).}
  \label{fig:bounds_sf}
\end{figure}

Theorem~\ref{thm:opt1} demonstrated that we require
\[
\Rnew(\Rreuse) \ge H(Y|U)+\Rreuse \mbox{.}
\]
The easily achieved corner points discussed
previously and a few simple bounds are shown in Fig.~\ref{fig:bounds_sf}.
The bounds, marked by dotted lines, are as follows:
\begin{itemize}
  \item[(a)] The lossless source coding theorem applied to $Y$ alone gives
        $\Rnew^*(\Rreuse) \geq H(Y)$.
  \item[(b)] A trivial lower bound from the construction is
        $\Rnew^*(\Rreuse) \geq \Rreuse$.
  \item[(c)] Since one could encode $Y_1^n$ without trying to take advantage
        of the $n\Rreuse$ symbols already available, $\Rnew^*(\Rreuse) \le \Rreuse+H(Y)$.
\end{itemize}

\subsection{Convexity of Regime}

In evaluating the properties of $\Rnew^*(\Rreuse)$ further,
let $W$ be a minimal sufficient statistic of $X$ for $Y$.
Intuitively, if $\Rreuse$ is large enough that one can encode $W$ in the
shared segment $U_1^{n\Rreuse}$, it is efficient to do so.
Thus we obtain regimes based on whether $\Rreuse$ is larger than $H(W)$.

For the regime of $\Rreuse \geq H(W)$, the boundary of the region is linear.
\begin{thm}
\label{thm:lm_regime2}
Consider the problem of \eqref{eqn:opt1}. Let $W$ be a minimal sufficient
statistic of $X$ for $Y$. For $\Rreuse > H(W)$, the solution is
given by:
\begin{equation}
\label{eqn:thm_highJ}
\Rupdate^*(\Rreuse) = \Rnew^*(\Rreuse) - \Rreuse =  H(Y|W) \mbox{.}
\end{equation}
\end{thm}
\begin{IEEEproof}
By definition, a sufficient statistic contains all information
in $X$ about $Y$.  Therefore any rate beyond the rate required
to transmit the sufficient statistic is not useful.
Beyond $H(W)$, the solution is linear.
\end{IEEEproof}
A rearrangement of \eqref{eqn:thm_highJ} is
\[
\Rnew^*(\Rreuse) =  H(Y,W) + [\Rreuse - H(W)]\mbox{.}
\]
This is used to draw the portion of the boundary determined by
Theorem~\ref{thm:lm_regime2} with a solid line in Fig.~\ref{fig:bounds_sf}.

For the regime of $\Rreuse < H(W)$, we have not determined the boundary
but we can show that $\Rnew^*(\Rreuse)$ is convex.
\begin{thm}
\label{thm:lm_regime1}
Consider the problem of \eqref{eqn:opt1}. Let $W$ be a minimal
sufficient
statistic of $X$ for $Y$. For $\Rreuse < H(W)$, the solution
$\Rnew^*(\Rreuse)$ is convex.
\end{thm}
\begin{IEEEproof}
Follows from the convexity of conditional entropy, by mixing
possible distributions $U$.
\end{IEEEproof}

The convexity from Theorem~\ref{thm:lm_regime1} and the unit slope of
$\Rnew^*(\Rreuse)$ for $\Rreuse > H(W)$ from Theorem~\ref{thm:lm_regime2}
yield the following theorem by contradiction.
An alternative proof is given in Appendix~\ref{app:proof}.
\begin{thm}
\label{thm:slopeBounds}
The slope of $\Rnew^*(\Rreuse)$ is bounded below and above:
\[
0 \leq \frac{d}{d\Rreuse} \Rnew^*(\Rreuse) \leq 1 \mbox{.}
\]
\end{thm}
The following are extremal cases of the theorem:
\begin{itemize}
  \item When $X$ and $Y$ are independent, $\Rnew^*(\Rreuse) = \Rreuse + H(Y)$ and so $\tfrac{d}{d\Rreuse}L^*(\Rreuse) = 1$
  \item When $X = Y$, $\Rnew^*(\Rreuse) = H(Y)$ for any $\Rreuse$, and so $\tfrac{d}{d\Rreuse}\Rnew^*(\Rreuse) = 0$.
\end{itemize}

\section{Connections}
\label{sec:connections}

An alternate method of further analyzing the rate region for fixed reuse
is to make connections with solved problems in the literature.  
A source coding problem intimately related to 
the G\'{a}cs--K\"{o}rner common information provides a partial converse.  

A seemingly related problem solved by Vasudevan and Perron \cite{VasudevanP2007}
does not provide too much further insight into our rate region.
Relating their problem statement to our 
problem statement requires the rate $\Robsolete$ in our problem setup to be 
set to $0$ and the decoder for $Y$ to decode both $(\hat X, \hat Y)$.

\subsection{Relation to G{\'{a}}cs--K{\"{o}}rner Common Information}

The G{\'{a}}cs--K{\"{o}}rner common information \cite{GacsK1973},
helps characterize the rate region.  It also 
arises in lossless coding with coded side information \cite{AhlswedeK1975,Wyner1975b,MarcoE2009}.

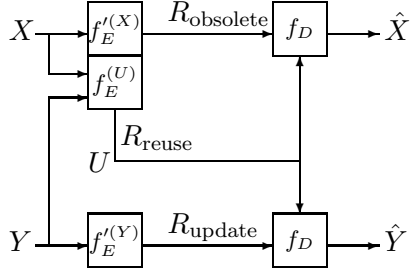
\begin{figure}
  \centering

  \setlength{\unitlength}{10pt}
  \begin{picture}(14.0,12.0)(0.0,1.4)

  \put(0.0,10.0){$X$}
  \put(0.0,2.0){$Y$}
  \put(1.0,10.2){\vector(1,0){2.0}}
  \put(1.0,2.2){\vector(1,0){2.0}}
  \put(1.5,10.2){\line(0,-1){1.5}}
  \put(1.5,2.2){\line(0,1){5.6}}
  \put(1.5,8.7){\vector(1,0){1.5}}
  \put(1.5,7.8){\vector(1,0){1.5}}
  \put(3.0,9.4){\framebox(2.0,2.0){\footnotesize $f_E^{\prime(X)}$}}
  \put(3.0,7.4){\framebox(2.0,2.0){\footnotesize $f_E^{(U)}$}}
  \put(3.0,1.4){\framebox(2.0,2.0){\footnotesize $f_E^{\prime(Y)}$}}
  \put(5.0,10.2){\vector(1,0){5.0}}
  \put(5.0,2.2){\vector(1,0){5.0}}
  \put(6.0,10.7){$\Robsolete$}
  \put(6.0,2.7){$\Rupdate$}
  \put(10.0,9.4){\framebox(2.0,2.0){\footnotesize $f_D$}}
  \put(10.0,1.4){\framebox(2.0,2.0){\footnotesize $f_D$}}
  \put(12.0,10.2){\vector(1,0){2.0}}
  \put(12.0,2.2){\vector(1,0){2.0}}
  \put(14.2,10.0){$\hat{X}$}
  \put(14.2,2.0){$\hat{Y}$}
  \put(4.0,7.4){\line(0,-1){2.0}}
  \put(3.0,5.0){$U$}
  \put(4.2,6.0){$\Rreuse$}
  \put(4.0,5.4){\line(1,0){7.0}}
  \put(11.0,5.4){\vector(0,1){4.0}}
  \put(11.0,5.4){\vector(0,-1){2.0}}
  \end{picture}
  \caption{Block diagram for a source network, \cite[Fig.~P.28 on p.~403]{CsiszarK1997}.}   
  \label{fig:commoninformation}
\end{figure}

\begin{defn}
For random variables $X$ and $Y$, let $U = f(X) = g(Y)$ where $f$ is a function 
of $X$ and $g$ is a function of $Y$ such that $f(X) = g(Y)$ almost surely and the
number of values taken by $f$ (or $g$) with positive probability is the largest
possible.  Then the \emph{G{\'{a}}cs--K{\"{o}}rner common information}, denoted $C(X;Y)$,
is $H(U)$.
\end{defn}
\begin{defn}
The joint distribution $p(x,y)$ is \emph{indecomposable} if there are no
functions $f$ and $g$ each with respect to the domain $\mathcal{W}$ so that
$\Pr[f(X) = g(Y)] = 1$, and $f(X)$ takes at least two values with non-zero probability.
\end{defn}
\begin{lem}
\label{lem:indec}
Common information $C(X;Y) = 0$ if $X$ and $Y$ have an indecomposable distribution.
\end{lem}
\begin{IEEEproof}
See \cite{GacsK1973}.
\end{IEEEproof}
\begin{lem}
\label{lem:common_info}
Consider the source network \cite[Fig.~P.28 on p.~403]{CsiszarK1997}, redrawn as Fig.~\ref{fig:commoninformation}.
The largest $\Rreuse$ for which the rate triple $(\Rreuse, \Robsolete = H(X) - \Rreuse, \Rupdate = H(Y) - \Rreuse)$ is achievable (with Shannon reliability)
is $\Rreuse = C(X;Y)$.
\end{lem}
\begin{IEEEproof}
See \cite[P28 on p.~404]{CsiszarK1997}.
\end{IEEEproof}

\begin{cor}
\label{cor:common_info_e}
Consider the source network in Fig.~\ref{fig:commoninformation_extended}.
The largest $\Rreuse$ for which the rate triple $(\Rreuse, \Robsolete = H(X) - \Rreuse, \Rupdate = H(Y) - \Rreuse)$ is achievable (with Shannon reliability)
is $\Rreuse = C(X;Y)$.
\end{cor}
\begin{IEEEproof}
Follows from Lemma~\ref{lem:common_info} and the Markov relation $U \leftrightarrow X \leftrightarrow Y$,
so additional knowledge of $U$ provides no benefit to $f_E^{\prime(Y)}$.
\end{IEEEproof}
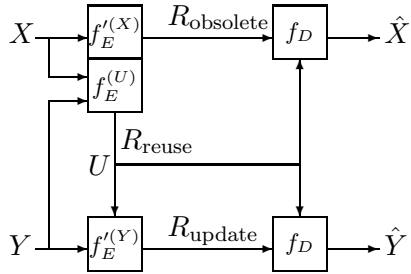
\begin{figure}
  \centering
  \setlength{\unitlength}{10pt}
  \begin{picture}(14.0,12.0)(0.0,1.4)
  \put(0.0,10.0){$X$}
  \put(0.0,2.0){$Y$}
  \put(1.0,10.2){\vector(1,0){2.0}}
  \put(1.0,2.2){\vector(1,0){2.0}}
  \put(1.5,10.2){\line(0,-1){1.5}}
  \put(1.5,2.2){\line(0,1){5.6}}
  \put(1.5,8.7){\vector(1,0){1.5}}
  \put(1.5,7.8){\vector(1,0){1.5}}
  \put(3.0,9.4){\framebox(2.0,2.0){\footnotesize $f_E^{\prime(X)}$}}
  \put(3.0,7.4){\framebox(2.0,2.0){\footnotesize $f_E^{(U)}$}}  
  \put(3.0,1.4){\framebox(2.0,2.0){\footnotesize $f_E^{\prime(Y)}$}}
  \put(5.0,10.2){\vector(1,0){5.0}}
  \put(5.0,2.2){\vector(1,0){5.0}}
  \put(6.0,10.7){$\Robsolete$}
  \put(6.0,2.7){$\Rupdate$}
  \put(10.0,9.4){\framebox(2.0,2.0){\footnotesize $f_D$}}
  \put(10.0,1.4){\framebox(2.0,2.0){\footnotesize $f_D$}}
  \put(12.0,10.2){\vector(1,0){2.0}}
  \put(12.0,2.2){\vector(1,0){2.0}}
  \put(14.2,10.0){$\hat{X}$}
  \put(14.2,2.0){$\hat{Y}$}
  \put(4.0,7.4){\vector(0,-1){4.0}}
  \put(3.0,5.0){$U$}
  \put(4.2,6.0){$\Rreuse$}
  \put(4.0,5.4){\line(1,0){7.0}}
  \put(11.0,5.4){\vector(0,1){4.0}}
  \put(11.0,5.4){\vector(0,-1){2.0}}
  \end{picture}
  \caption{Block diagram for another source network.}   
  \label{fig:commoninformation_extended}
\end{figure}

Having reviewed extant results on the G{\'{a}}cs--K{\"{o}}rner common information
and extended them slightly, we use them to characterize the malleable coding problem.

\begin{figure}
  \centering

  \setlength{\unitlength}{10pt}
  \begin{picture}(14.0,12.0)(0.0,1.4)

  \put(0.0,10.0){$X$}
  \put(0.0,1.0){$Y$}
  \put(1.0,10.2){\vector(1,0){2.0}}
  \put(1.0,1.2){\vector(1,0){2.0}}
  \put(1.5,10.2){\line(0,-1){1.5}}
  \put(1.5,8.7){\vector(1,0){1.5}}
  \put(3.0,9.4){\framebox(2.0,2.0){\footnotesize $f_E^{\prime(X)}$}}
  \put(3.0,7.4){\framebox(2.0,2.0){\footnotesize $f_E^{(U)}$}}  
  \put(3.0,0.4){\framebox(2.0,2.0){\footnotesize $f_E^{\prime(Y)}$}}
  \put(5.0,10.2){\vector(1,0){5.0}}
  \put(5.0,1.2){\vector(1,0){5.0}}
  \put(6.0,10.7){$\Robsolete$}
  \put(6.0,1.7){$\Rupdate$}
  \put(10.0,9.4){\framebox(2.0,2.0){\footnotesize $f_D$}}
  \put(10.0,0.4){\framebox(2.0,2.0){\footnotesize $f_D$}}
  \put(12.0,10.2){\vector(1,0){2.0}}
  \put(12.0,1.2){\vector(1,0){2.0}}
  \put(14.2,10.0){$\hat{X}$}
  \put(14.2,1.0){$\hat{Y}$}
  \put(4.0,7.4){\vector(0,-1){2.2}}
  \put(3.0,4.2){\framebox(2.0,1.0){$\quad$}}
  \put(4.0,4.2){\vector(0,-1){1.8}}
  \put(3.0,2.9){$\hat{U}$}

  \put(3.0,5.8){$U$}
  \put(4.2,6.0){$\Rreuse$}
  \put(4.0,5.8){\line(1,0){7.0}}
  \put(11.0,5.4){\vector(0,1){4.0}}
  \put(11.0,5.4){\vector(0,-1){3.0}}

  \end{picture}

  \caption{Block diagram for malleable coding with fixed reuse in extended form.}   
  \label{fig:commoninformation_enhanced_cp}
\end{figure}
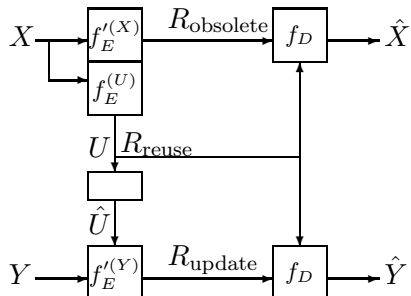

\begin{thm}
The rate triple $(\Rreuse = C(X;Y), \Robsolete = H(X)-C(X;Y), \Rupdate = H(Y)-C(X;Y))$ provides a partial converse to
the rate triple $\mathcal{M}$ for malleable coding.
\end{thm}
\begin{IEEEproof}
Using a block-diagrammatic information flow representation, a greater number of lines and a smaller number of noisy channel boxes
both signify more extensive information patterns.
The source network in Fig.~\ref{fig:commoninformation_extended} has a more extensive information
pattern than in the malleable coding problem (see Fig.~\ref{fig:commoninformation_enhanced_cp}).
Thus, the result follows from Corollary~\ref{cor:common_info_e}.
\end{IEEEproof}

The interpretation of this result is that if want $\Rold = H(X)$
and $\Rnew=H(Y)$ for the malleable coding problem, then $\Rupdate$ must be large: $\Rupdate\ge H(Y)-C(X;Y)$.
In general $C(X;Y)=0$ by Lemma~\ref{lem:indec}, so in this case the stored symbols
cannot be reused at all, asymptotically.

\section{Discussions and Closing Remarks}
\label{sec:final}

Phrased in the language of waste avoidance and resource recovery:
classical Shannon theory shows how to optimally \emph{reduce};
we have here studied \emph{reuse} and in \cite{VarshneyKG2009} studied \emph{recycling},
and we have found these goals to be fundamentally in tension.

We have formulated an information-theoretic problem motivated by
the transmission of data to edit the compressed version of a
document after it has been updated. 
Any technique akin to optimally compressing the difference
between the documents would require the receiver to uncompress,
apply the changes, and recompress.
We instead require reuse of a fixed portion of the
compressed version of the original document;
this segment cut from the compressed version of the original document
is pasted into the compressed version of the new document.
This requirement creates a trade-off between the amount of reuse and
the efficiency in compressing the new document.
Theorem~\ref{thm:opt1} provides a complete characterization as a
single-letter information-theoretic optimization.

By establishing a relationship with the
G{\'{a}}cs--K{\"{o}}rner common information problem,
we see that if the original
and modified sources have an indecomposable joint distribution and
are required to be coded close to their entropies,
then the reused fraction must asymptotically be negligible.

\section*{Acknowledgment}
The authors thanks R.\ Johari, S.~K.\ Mitter, \.{I}.~E.\ Telatar, and
V.\ Tarokh for discussions and D.\ Marco for providing a preprint 
of \cite{MarcoE2009}.

\begin{appendices}
\section{Continuity Lemma}
According to \cite[Theorem 3.2.i]{Topsoe2001}, the entropy  function
is continuous in total variation over finite alphabets, cf.~\cite[Lemma~6]{EswaranSSG2010}.
We use this.
\begin{lem}
\label{lem:continuity}
$H(Y|\hat{U}) - H(Y|U) \to 0$ as $n\to \infty$
\end{lem}
\begin{IEEEproof}
First note that $H(Y_1^n|U_1^n) = nH(Y|U)$ and $H(Y_1^n|\hat{U}_1^n) = nH(Y|\hat{U})$
by memorylessness.  Therefore
\[
H(Y|\hat{U}) - H(Y|U) = \tfrac{1}{n}\left[H(Y_1^n|\hat{U}_1^n) - H(Y_1^n|U_1^n)\right] \mbox{.}
\]

Let us proceed with considering $H(Y_1^n|\hat{U}_1^n) - H(Y_1^n|U_1^n)$.   
We know that $\Pr[U_1^n \neq \hat{U}_1^n] \le
\epsilon$, $\epsilon \to 0$ as $n\to \infty$, by the a.a.s.\
lossless coding.  We also know that the Markov condition
$\hat{U}_1^n \leftrightarrow U_1^n \leftrightarrow Y_1^n$ holds.

It follows from the Markov relation and the error probability bound
that we can bound the variational distance
\[
\| p_{Y_1^n|\hat{U}_1^n} - p_{Y_1^n|U_1^n} \|_1 \le K_1(\epsilon,|\mathcal{U}|)
\]
where $K_1$ is a fixed constant that depends on the error
probability $\epsilon$ and alphabet size $|\mathcal{U}|$, since
$p_{Y_1^n|\hat{U}_1^n} = p_{Y_1^n|U_1^n}p_{U_1^n|\hat{U}_1^n}$ by
Markovianity, so $p_{Y_1^n|\hat{U}_1^n} - p_{Y_1^n|U_1^n} =
(-\vec{1} + p_{U_1^n|\hat{U}_1^n})p_{Y_1^n|U_1^n}$ and $-\vec{1} +
p_{U_1^n|\hat{U}_1^n}$  is small by the error bound.

Now since entropy is continuous in variational distance for finite
alphabets \cite[Theorem 3.2.i]{Topsoe2001}, the result follows.
\end{IEEEproof}

\section{Alternate Proof of Theorem~\ref{thm:slopeBounds}}
\label{app:proof}

\emph{Proof of upper bound:}
Let $\Rreuse^{(1)} > \Rreuse^{(2)}$ be any two values of $\Rreuse$.  Let $V_1$ and $V_2$
be the corresponding auxiliary random variables $U$ that solve
the optimization problem \eqref{eqn:opt1}.  Then by the
successive refinability of lossless coding, it
follows that $V_1$ and $V_2$ will satisfy the Markov chain
$V_2 \leftrightarrow V_1 \leftrightarrow X \leftrightarrow Y$.

By the data processing inequality,
\begin{align*}
I(Y;V_2) &\le I(Y;V_1) \\ \notag
H(V_1|Y) - H(V_2|Y) &\le H(V_1) - H(V_2) \mbox{.}
\end{align*}
By definition,
\begin{eqnarray*}
\lefteqn{\Rnew^*(\Rreuse^{(1)} ) - \Rnew^*(\Rreuse^{(2)} )} \\
 & = & H(Y|V_1) + H(V_1) - H(Y|V_2) - H(V_2)  \\
 & = & H(V_1|Y) - H(V_2|Y) \mbox{.}
\end{eqnarray*}
Therefore,
\[
\Rnew^*(\Rreuse^{(1)} ) - \Rnew^*(\Rreuse^{(2)} ) \le H(V_1) - H(V_2) = \Rreuse^{(1)}  - \Rreuse^{(2)} 
\]
which implies
\[
\frac{\Rnew^*(\Rreuse^{(1)} ) - \Rnew^*(\Rreuse^{(2)} )}{\Rreuse^{(1)} - \Rreuse^{(2)}} \le 1 \mbox{.}
\]

\emph{Proof of lower bound:} We want to show that $H(V_1|Y) -
H(V_2|Y) \ge 0$. This property may be verified using Yeung's ITIP
\cite{Yeung2002} after invoking the Markov chain $V_2
\leftrightarrow V_1 \leftrightarrow X \leftrightarrow Y$ and the
subrandomness conditions, $H(V_1|X) = H(V_2|X) = 0$.

\end{appendices}

\bibliographystyle{IEEEtran} 
\bibliography{abrv,conf_abrv,lrv_lib}
\end{document}